\documentstyle[axodraw,12pt]{article}

\newcommand{\be}{\begin{eqnarray}}
\newcommand{\ee}{\end{eqnarray}}
\newcommand{\non}{\nonumber\\}

\newcommand{\inline}[1]{\noalign{\hbox{#1}}}

\newcommand{\pref}[1]{(\ref{#1})}
\advance\textheight by 1in
\advance\voffset by -0.5in

\title{Minding one's P's \& Q's : from the one loop effective action in 
quantum field theory 
to classical transport theory}
\author{Jamal Jalilian-Marian\\
{\small \it Physics Department, University of Arizona, Tucson, AZ 85721.}\\
Sangyong Jeon\\
{\small \it Nuclear Science Division,}\\
{\small \it Lawrence Berkeley National Laboratory, Berkeley, CA 94720.}\\
Raju Venugopalan\\
{\small \it Physics Department, 
Brookhaven National Laboratory, Upton, NY 11973.}\\
Jens Wirstam\\
{\small \it Institute for Theoretical Physics, University of Stockholm,}\\ 
{\small \it Box 6730, S-113 85, Stockholm, Sweden.}\\
   }

\begin{document}
\maketitle

\begin{center}
{\bf Abstract}\\
\end{center}

The one loop effective action in quantum field theory can be expressed
as a quantum mechanical path integral over world lines, with internal
symmetries represented by Grassmanian variables. In this paper, we
develop a real time, many body, world line formalism for the one loop
effective action. In particular, we study hot QCD and obtain the
classical transport equations which, as Litim and Manuel have shown,
reduce in the appropriate limit to the non--Abelian
Boltzmann--Langevin equation first obtained by B\"{o}deker.  In the
Vlasov limit, the classical kinetic equations are those that
correspond to the hard thermal loop effective action.  We also
discuss the imaginary time world line formalism for a hot $\phi^4$
theory, and elucidate its relation to classical transport theory.

\vfill \eject

 \section{Introduction}

 In classical kinetic theory, a covariant formalism can be obtained in
 terms of phase space averages over the trajectories of particle world
 lines~\cite{Israel}. In quantum field theory, there are many
 instances at finite temperature and density where classical ideas are
 relevant, and where a classical kinetic picture would be
 useful. However, it is not immediately apparent how one recovers
 the classical world line picture directly from quantum field
 theory. This is especially problematic in theories with internal symmetries.

 Fortunately, in the last decade, there has been a considerable body
 of work relating the one loop effective action in quantum field theory
 to quantum
 mechanical path integrals over point particle Lagrangians\footnote{The
 connection between fields and particles is of course relatively
 ancient. It goes back to the works of Feynman and
 Schwinger~\cite{FeynSchwin}.  Also, Polyakov in his
 book~\cite{Polyakov} demonstrates the relation between particle Green
 functions and the n-point Feynman amplitudes of quantum field theory.}. 
For a survey of recent
 developments, see the review by Schubert~\cite{Schubert}.
 These recent developments follow from the key insight by Berezin and 
 Marinov~\cite{BerMarinov} that internal symmetries such as color and 
 spin had classical analogues in terms of Grassmanian variables. These 
 obeyed classical commutation relations which, when quantized, gave the 
 usual commutation relations for spin and color. 
 
 Brink, DiVecchia, and Howe used  
 these Grassmanian variables to
 construct a classical Lagrangian for spinning world lines in an
 Abelian background field~\cite{BriDiHo}.  Subsequently, Balachandran
 et al.~\cite{BalSkag} and Barducci et al.~\cite{Bard} wrote down the
 following Lagrangian for a classical colored particle in a
 non-Abelian background field\footnote{They also considered the case
 of a classical colored, spinning, particle in a non-Abelian
 background field. We will not discuss this case here.}:
 \be
 L = -m \sqrt{{\dot{x}}_\mu {\dot{x}}^\mu}  + i \lambda_a^\dagger D_{ab} 
 \lambda_b \, .
 \label{BSBL}
 \ee
 Here the $\lambda_a (\tau)$ 
 with $a=1,\cdots,N$ are Grassmanian dynamical variables, and 
 $D_{ab}\lambda_b 
 = {\dot{\lambda}}_a + ig{\dot{x}}^\mu A_\mu^\alpha T_{ab}^\alpha \lambda_b$. 
 The variables $T_{ab}^\alpha$'s are $N\times N$ matrices in an irreducible
 representation of the Lie algebra of the group.  The Euler-Lagrange
 equations of motion are deduced in the usual way from the above
 Lagrangian. It was shown by Balachandran et al.\ and by Barducci 
 et al.\ that these equations are precisely the equations written down 
 nearly thirty years ago by S.K.~Wong~\cite{SKWong}.

 The connection of the work on point particle Lagrangians to quantum
 field theory was first made by Strassler~\cite{Strassler}. He showed
 that the one loop effective action in quantum field theory could be
 expressed in terms of a quantum mechanical path integral over a point
 particle Lagrangian. For an Abelian gauge theory, Strassler
 showed that this Lagrangian was identical to the one written down
 previously by Brink, DiVecchia, and Howe~\cite{Strassler}. For
 non-Abelian gauge theories, Strassler wrote down an expression, but
 it did not have a Lagrangian interpretation since the trace over
 color--the Wilson loop--was kept explicitly.  The corresponding world
 line Lagrangian was first obtained by D'Hoker and Gagn\'{e} using a
 coherent state formalism~\cite{DHokerGagne}.
 Remarkably, as observed by Pisarski and Tytgat~\cite{Roblectures},  
 the point particle Lagrangian D'Hoker and
 Gagn\'{e} obtained by integrating out fermions 
 coupled to a vector background field, coincides with the  Lagrangian
 in Eq.~\pref{BSBL}.

 Pisarski and Tytgat have noted that since one can write the one loop effective
 action as a path integral over world lines, it is a possible
 explanation of the apparently mysterious result that classical
 kinetic theory in terms of the single particle distributions
 $f(x,p,Q)$ gives rise to the well known HTL effective action for
 finite temperature gauge theories~\cite{Roblectures}. In this work, we
 extend the previous work on the vacuum world line formalism, and develop 
 a real time many body formalism, which may be applied to a wide range of 
 many body problems at finite temperature and density. In particular, we 
 will focus our attention here
 on the apparently mysterious results that were discussed  
 in Ref.~\cite{Roblectures}, 
 and show how they may be understood in the many body world 
 line formalism. Before we can be more explicit, we need therefore 
 to discuss the status of recent work in finite 
 temperature gauge theories.      

 At very high temperatures, the physics of soft modes in gauge
 theories may be described by effective field theories~\cite{Braaten}.
 Integrating out modes with momenta $p\sim T$, one obtains an
 effective theory at the scale $p\sim gT$, characterized by a Debye
 screening mass $m_D^2 \propto g^2 T^2$.  The gauge invariant
 effective action describing these modes is often referred to as the
 Hard Thermal Loop (HTL) effective
 action~\cite{BraatenPisarski,Othaut,Nair}.  Recently, there has been much
 progress in describing very soft magnetic modes at the scale $p\sim
 g^2 T$ by systematically integrating out modes at the scales $T$ and
 $gT$~\cite{Bodeker,MooreRummu,LitimManuel,Eddie,ASY}.

 It was first shown by Blaizot and Iancu~\cite{BlaiIancu1} that the
 non-local hard thermal loop effective action could be derived in a
 local kinetic approach by systematically truncating the
 Schwinger-Dyson equations. The kinetic approach was very useful since it 
 lead to an intuitive picture in which soft modes $p\sim gT$  satisfy 
 classical equations of motion and are coupled to hard modes $p\sim T$. 
 These hard modes in turn satisfy the collisionless Boltzmann equation. 

 This Schwinger-Dyson approach is a perfectly reasonable way to proceed. 
 However, since the approximations are made at the level of the operator 
 equations of motion, it can become cumbersome. A local effective action 
 was developed by Iancu~\cite{Eddie2}, which is more useful, and indeed 
 has been the basis of some recent work~\cite{MooreRummu,Moore}. 

 Some time later, Kelly et al.~\cite{Kellyetal} wrote down an alternative  
 transport theory for classical colored particles. These particles obey 
 the classical equations of motion of a spinless colored particle, coupled 
 to a non-Abelian background field $A_\mu^a$, 
 that were first written down by Wong~\cite{SKWong}.
 
 In the picture of 
 Kelly et al., the single particle distributions $f(x,p,Q)$ 
 are defined for an extended phase space of classical colored charges. (In 
 this regard, their approach is similar to the earlier work of 
 Heinz~\cite{Heinz1} and collaborators 
 on a classical transport theory for colored charges.) 
 They showed that this distribution obeys the collisionless 
 Boltzmann equation.  Making suitable approximations 
 to their kinetic equations, they showed 
 that they also recovered the HTL effective action.

 The classical transport theory of Kelly et al.\ is very simple and
 elegant.  However, the formalism is {\it ad hoc}, and it was not
 clear whether such a formalism could be ``derived'' in some
 systematic approximation.  It was also not clear whether the
 distributions $f$ were distributions in the statistical mechanics
 sense, since no phase space averaging had been performed in the
 derivation. Finally, as pointed out by Blaizot and
 Iancu~\cite{BlaiIancu2}, it was not clear why color could be treated
 as a classical degree of freedom.
 
 The problematic issue of statistical averaging in the work of Kelly et 
 al.~was clarified recently in the work of 
 Litim and Manuel~\cite{LitimManuel}.
 In this work, we show how, in the real time many body world line formalism, 
 the classical transport theory of Kelly et al., as improved by Litim and 
 Manuel, may be obtained from the one loop effective action in QCD. We 
 obtain the same set of transport equations they do. Since our results are 
 derived from the QCD one loop effective action, one can, in principle, go 
 further. We will not attempt to do so in this paper, but save that effort 
 for later work.

 This paper is organized as follows. In section 2, we summarize the
 world line formalism for the one loop effective action in QCD at zero
 temperature. In section 3, we discuss the point particle Lagrangian
 that results from the one loop effective action. We show that Wong's
 equations follow from this Lagrangian. We comment on the classical
 commutation relation among the various dynamical variables and their
 quantum counterparts. Next, in section 4, a many body world line
 formalism is developed, and applied to the one loop effective action.
 In the following section, it is shown that the saddle point of this
 effective action gives the finite temperature classical transport
 theory of Litim and Manuel. As shown by these authors, their results 
 agree with those obtained previously by B\"{o}deker~\cite{Bodeker} 
 for soft modes in hot gauge theories. The final section explores the
 implications of our result, and discusses further avenues of
 research. In appendix A, we work out, in the imaginary time approach,
 a simple example illustrating the finite temperature world line
 formalism. We also point out, in this context, the connections to classical 
 transport theory.

 \section{The one loop effective action and the world line formalism 
 in quantum field theory}

 In recent years, the world line formalism has become popular as a
 technique to compute one loop scattering amplitudes 
 in gauge theories. 
 Initial interest in such techniques followed from the work of
 Bern and Kosower~\cite{BernKosower}, who derived
 rules from string theory for computing one loop scattering amplitudes
 in gauge theories. Strassler showed that similar rules
 could be obtained directly in gauge theories by writing the one loop
 effective action, in background field gauge, as a one dimensional
 quantum mechanical path integral over the ``world line'' of a point
 particle in an external background field~\cite{Strassler}.
  
 We consider, with Strassler as a guide, the example of the 
 Lagrangian of a massless
 scalar field coupled to a background Abelian gauge field $A_\mu(x)$,
 \be
 {\cal L} = \Phi^\dagger D^2 \Phi \, ,
 \label{scalLag}
 \ee
 where $D_\mu = \partial_\mu-igA_\mu$. The one loop effective action
 is 
 \be
 \Gamma[A] = -\log\left[{\rm det}(-D^2)\right] \equiv 
 - {\rm Tr}\left(\log(-D^2)\right)\, .
 \label{sc1lp}
 \ee
 (The general case of Dirac 
 spinors coupled to scalars, vectors, axial scalars and 
 vectors, and tensors, has been considered by several authors\footnote{ 
 for  a compact review, see the talk by C. Schubert~\cite{HellSchool}.}.)

 We then use the trick 
 \be
 \log(\sigma) = 
 \int_1^\sigma \,{dy\over y}\equiv \int_1^\sigma \,dy \int_0^\infty 
 dt\, e^{-y t} 
 = -\int_0^\infty\, {dt \over t} \left(e^{-\sigma t}-e^{-t}\right) \, , 
 \label{trick1}
 \ee
 to re-write Eq.~\pref{sc1lp} as
 \be
 \Gamma[A] &=& \int_0^\infty\, {dT\over T} \int \, {d^4 p \over {(2\pi)^4}} 
\, \langle p|\exp{\left(-{1\over 2} {\varepsilon}T\left(p+gA(x)\right)^2
\right)}|p\rangle \nonumber \\
 &=& \int_0^\infty \, {dT\over T}\, {\cal N}\,
 \int {\cal D}x\, {\cal P} \exp \left[
 -\int_0^T d\tau\, \left( {1\over 2{\varepsilon}}{\dot{x}}^2 +ig 
 A[x(\tau)]\cdot \dot{x}\right)\right] 
 \label{trick2}
 \ee
 where ${\cal P}$ denotes the path-ordering.

 Above, ${\varepsilon}$ is the einbein (the square root of the one 
 dimensional metric--taken here to be an arbitrary constant) and 
 \be
 {\cal N} = \int {\cal D} p\, \exp\left( -{1\over 2} \int_0^T d\tau\, 
 {\varepsilon}\, p^2\right) \, ,\nonumber 
 \ee
 is a normalization constant. Analytically continuing this expression to 
 Minkowski space-time and letting ${\varepsilon}\longrightarrow 
 -{\varepsilon}$, one obtains in the general case of $A_\mu$ being a 
 matrix in the group representation $R$ of the scalar field
 \be
 \Gamma[A] &=& \int_0^\infty\, {dT\over T}\, {\cal N}\,\int {\cal D}x\, 
 {\rm Tr}_R\,
 {\cal P} 
 \exp \left[
 i\int_0^T d\tau\, \left( {1\over 2{\varepsilon}}{\dot{x}}^2 -ig 
 A[x(\tau)]\cdot \dot{x}\right)\right] \, 
 \label{Strassler}
\non
 \ee
 where ${\rm Tr}_R$ denotes the trace over the internal states.
 This expression can be factorized into a product of exponentials--the $A$ 
 dependent piece being a Wilson loop. It is therefore 
 the expectation value of a Wilson loop of the background field, in a 
 particular ensemble of loops. 

 Strassler showed that one particle irreducible (1PI) 
 Feynman diagrams, with $N$ 
 external background gluons and a scalar loop, could be constructed by 
 expanding the above expression for the one loop effective action to order 
 $g^N$. Expanding the gauge field in a set of plane waves with definite 
 momentum, polarization, and color charge, he re-wrote the above 1PI-action 
 as factorized products of the traces over color and the $N$ proper 
 time integrals (and permutations thereoff). He then considered several 
 examples--and computed, for instance, the gluon vacuum polarization in a 
 theory with Dirac fermions and complex adjoint scalars.  

 Our objective here is slightly different. We wish to write 
 Eq.~\pref{Strassler} in terms of a point particle Lagrangian of scalar 
 particles which carry internal symmetries and are coupled to an  
 external field. (The extension to this work to fermions is straightforward 
 but will not be discussed here.) Towards this end, we shall here follow the 
 later work of D'Hoker and Gagn\'{e} (see also the related work of 
 Mondrag\'{o}n et al.~\cite{Myriam}). D'Hoker and Gagn\'{e} proved the 
 following identity for an $n\times n$ Hermitean matrix $M(\tau)$, 
 \be
 {\rm Tr}\,{\cal P} \exp\left( i\int_0^T\,d\tau\, M(\tau)\right) &=& 
\left({\pi\over T}\right)^n \sum_\phi \int_{AP} {\cal D}\lambda^\dagger 
 {\cal D}\lambda \, e^{i\phi(\lambda^\dagger\lambda +{n\over 2} -1)}
 \nonumber \\
 &\times&
\exp\left(-\int_0^T\, d\tau [\lambda^\dagger\dot{\lambda} -i\lambda^\dagger 
 M\lambda]\right) \, .
 \label{DHoker}
 \ee
 Here $\phi = 2\pi k/n$, where $k=1,\cdots,n$. As noted in the introduction, 
 the $\lambda^\dagger$ and $\lambda$ are independent dynamical Grassmanian 
 variables (the subscript ``AP'' denotes anti--periodic boundary conditions)
 which act as eigenvalues of
 creation and annihilation Fermi operators, 
 respectively, generating arbitrary finite dimensional representations of 
 the $SU(N_c)$ symmetry group.
 These are discussed further in the following section.
 Also, the value of 
 $\lambda^\dagger \lambda$ can be evaluated at any arbitrary value of 
 $\tau$. 

 In the adjoint case, the matrix $M$ stands for the adjoint gauge field 
 $A^a T^a$, with 
 $n = N_c^2-1$. Substituting the above identity in Eq.~\pref{Strassler}, 
 we can write  the one loop effective action for 
 the Lagrangian in Eq.~\pref{scalLag} as, 
 \be
 \Gamma[A] &=& \int_0^\infty\, {dT\over T}\, {\cal N}\,\int {\cal D}x\, 
 \int_{AP} {\cal D}\lambda^\dagger\,{\cal D}\lambda\,
 {1\over n}\,\left({\pi\over T}\right)^n \sum_{k=1}^n\, 
 e^{2\pi ik(\lambda^\dagger\lambda +{n\over 2} -1)/n}\nonumber \\
 &\times& \exp\left(i\int_0^T\, d\tau\, {\cal L}_p (\tau)\right) \, ,
 \label{finscalact}
 \ee
 with 
 \be 
 {\cal L}_p (\tau) = {{\dot{x}}^2\over 2{\varepsilon}} + 
 \lambda^\dagger \dot{\lambda} + ig {\dot{x}}_\mu 
 \lambda^\dagger A^\mu \lambda \, .
 \label{pointLag}
 \ee
 The important point to note here~\cite{Roblectures} is that the above 
 Lagrangian, contained in the one loop effective action, is {\it precisely} 
 the point particle Lagrangian of Balachandran et al., and of Barducci 
 et al.\ in Eq.~\pref{BSBL}. This result is extremely suggestive, and 
 is indeed the starting point for our work. 
 The above result in Eqs.~\pref{finscalact}
 and \pref{pointLag} can be easily extended to the full QCD case--the 
 expression for $\Gamma[A]$ above will also include a path integral over 
 classical spins (see Ref.~\cite{DHokerGagne}). Again, this Lagrangian 
 is identical to the generalized Wong Lagrangian~\cite{BalSkag}.

 In the following section, we will discuss Wong's equations and show how 
 they follow from Eq.~\pref{pointLag}. 

 \section{Wong's equations and the world line Lagrangian} 

 In recent years, Wong's equations for classical charged particles
 interacting with classical non-Abelian gauge fields have received
 a considerable amount of attention in finite temperature
 applications~\cite{Kellyetal,Heinz1,HuMuller}.  Wong wrote down the following 
 set of equations~\cite{SKWong} 
 \be
 m {dx_I^\mu\over d\tau} & = & mv^\mu_I = p_I^\mu 
 \label{eq:xdot}
 \\
 {dp_I^\mu\over d\tau}
 & = &
 g v_\nu^I\, {\rm Tr}\, \left(Q_I\, F^{\mu\nu}(x_I)\right)
 \label{eq:pdot}
 \\
 \dot{Q_I} & = & -ig\left[Q_I, v_\mu^I A^\mu\right]
 \label{eq:Qdot}
 \\
 D^\nu F_{\nu\mu} & = &  j_\mu
 \label{eq:DF=j}
 \\ 
 \inline{and}
 j_\mu(x) 
 & = &
 g\sum_{I=1}^K
 \int d\tau\, Q_I(\tau)\, v_\mu^I(\tau)\, 
 \delta^{4}\left[x - x_I(\tau)\right] \, .
 \label{wongeq}
 \ee
 Above, $\tau$ is the proper time, $Q^I(\tau) = Q_a^I(\tau)\,T_a$ is the color
 charge of the $I$-th particle 
 in the adjoint representation, and $(D_\mu)^{ab} = \delta^{ab}
 \partial_\mu + gf^{abc}A_\mu^c$ is the usual covariant derivative. 
 The second equation is the generalization of the Lorentz force
 equation, the third equation describes color precession.
 The fourth equation is the Yang--Mills equation, which gives us 
 \be
 D_\mu j^\mu = 0 \,\,
 \label{eq:Dj=0}
 \;,
 \ee
 where $j^\mu$ is the net color current generated by the classical system of 
 $K$ particles. 
 A similar set of equations including the spin 
 was also obtained by Heinz \cite{Heinz:1984}
 starting from the Dirac equation for quarks.
 
 Balachandran et al. showed that the above set of equations can be obtained
 from the point particle Lagrangian in Eq.~\pref{pointLag}. We will summarize
 their results below. The action 
 \be
 S = -\int\, d^4x\, {1\over 4} F_{\mu\nu}^a F^{a,\mu\nu} + \int\, d\tau\, 
 {\cal L}_p (\tau) \, ,
 \label{pointAct}
 \ee
 is a) real upto a total time derivative, b) re-parametrization invariant, 
 under $\tau\rightarrow \tau^\prime$, and c) invariant under gauge 
 transformations in which {\it both} 
 $A_\mu^\alpha$ and $\lambda_a$ are transformed. Thus under an infinitesimal 
 gauge transformation $\theta^\alpha$,
 \be
 A_\mu^\alpha &\longrightarrow& (A_\mu^\alpha)^\theta 
 = A_\mu^\alpha - g f_{\alpha\beta\gamma} 
 \theta^\beta A_\mu^\gamma -\partial_\mu \theta^\alpha \,,\nonumber\\
 \lambda_a &\longrightarrow& (\lambda_a)^\theta = \lambda_a +ig\theta^\alpha 
 T_{ab}^\alpha \theta_b \, .
 \ee
 
 Redefining $\lambda^{\dagger} \rightarrow i\lambda^{\dagger}$, 
 one obtains the equations of motion
 \be
 \dot{\lambda}_a + ig{\dot{x}}_\mu A_\mu^\alpha T_{ab}^\alpha \lambda_b = 0
 \, , 
 \label{lambdaeq}
 \ee
 and 
 \be
 {\partial \over \partial\tau} \left({{\dot{x}}_\mu\over {\varepsilon}}
 \right) +g Q^\alpha F_{\mu\nu}^\alpha {\dot{x}}^\nu = 0\, ,
 \label{accel}
 \ee
 where 
 \be
 Q^\alpha = \lambda_a^\dagger T_{ab}^\alpha \lambda_b \, .
 \label{Qdef}
 \ee
 From the preceding equation, and from
 Eq.~\pref{lambdaeq}, we obtain the equation of 
 motion for $Q^\alpha$ in Eq.~\pref{eq:Qdot}.

 Finally, taking the functional derivative of Eq.~\pref{pointAct} with
 respect to $A^\mu (x)$, we obtain the Yang-Mills equations in
 Eq.~\pref{eq:DF=j}, with the current $j^\mu$ defined as in
 Eq.~\pref{wongeq}.

 In the Hamiltonian formalism, $x^\mu$, $\lambda_a$ and $\lambda_a^\dagger$ 
 are independent dynamical variables. Their momentum conjugates are 
 \be
 P_\mu &=& {{\dot{x}}_\mu\over {\varepsilon}}-gQ^\alpha A_\mu^\alpha \, .
 \nonumber \\
 P_a &=& i\lambda_a^\dagger \, . \nonumber \\
 P_a^\dagger &=& 0 \, .
 \ee
 respectively.
Due to the second class constraints, we have to introduce Dirac brackets. For two dynamical
variables $A$ and $B$ which are even elements of the Grassman algebra we have,
\be
 \left\{A,B\right\}_D = A\left( {\stackrel{\leftarrow}{\partial}\over 
 \partial x^\mu}{\partial \over \partial P^\mu} - 
 {\stackrel{\leftarrow}{\partial}\over 
 \partial P^\mu}{\partial \over \partial x^\mu}
-i{\stackrel{\leftarrow}{\partial}\over 
 \partial \lambda_a^\dagger}{\partial \over \partial \lambda_a}
-i{\stackrel{\leftarrow}{\partial}\over 
 \partial \lambda_a}{\partial \over \partial \lambda_a^{\dagger}} \right) B \, .
\label{DComm}
 \ee
This gives rise to the following Dirac brackets,
\be
 \left\{x^\mu,P_\nu\right\}_D &=& \delta_\nu^\mu \, . \nonumber \\
 \left\{\lambda_a,\lambda_b^\dagger\right\}_D &=& -i\delta_{ab} \nonumber \\
 \left\{ Q^\alpha,Q^\beta\right\}_D &=& f_{\alpha\beta\gamma}Q^\gamma \, .
 \ee

 Upon quantization, these give the usual Heisenberg commutation relation for 
 $P$ and $x$, and $\left\{\lambda_a,\lambda_b^\dagger\right\} = \delta_{ab}$.
 Also, $\left\{Q^\alpha,Q^\beta\right\}_D$ goes over into the usual commutator.
 As pointed out by Balachandran et al. and by Barducci et al., the quantized 
 $\lambda^\dagger$'s and $\lambda$'s act as Fermi creation and annihilation 
 operators, and generate various representations of the color group. The $n$ 
 creation and annihilation operators span the Clifford algebra $C_{2n}$. 
 The sum over $\phi$'s in Eq.~\pref{DHoker} may therefore be understood as 
 restricting the states generated by $\lambda$ and $\lambda^\dagger$ to a 
 particular irreducible representation of the group (see for instance 
 Eq.~(3.9) in Ref.~\cite{Bard}).

 We have dwelt at some length on the classical and quantum commutation
 relations among the dynamical variables because they will be useful
 later on in understanding the transition from the one loop effective
 action to classical transport theory.
 
 Before we end this section, we should mention that the generalization of 
 the action in Eq.~\pref{pointAct} to include classical spins was also 
 performed by Balachandran et al.,
 and by Barducci et al. The classical spins are also Grassmanian variables, 
 and upon quantization, go over into the Dirac $\gamma$-matrices. One 
 obtains the generalized Wong equations, including a generalized 
 Bargmann-Michel-Telegdi equation for the time evolution of the 
 Pauli-Lubanski spin four-vector. The generalized point particle Lagrangian 
 is again identical to the one that appears in the one loop effective action 
 for fermions in a classical background field.

 \section{The many body world line formalism}

 In this section, we will extend the formalism of section 2 to study a 
 many body system of world line scalars. This fills in the step where 
 one goes from a quantum field theoretic description in terms of ensemble 
 averages over fields, to the classical transport approach where one 
 takes phase space averages over world line trajectories. These averages 
 will of course correspond to 
 the multi-particle distributions of classical transport theory. 

 In general, for the path integral of a many particle system, one has
 an ensemble of initial and final 
conditions\footnote{For an excellent reference on path integrals for
 many particle systems, we recommend the book by Negele and
 Orland~\cite{NegeleOrland}.}(which do not necessarily
 coincide!). In addition, not all initial and final
 conditions are weighted equally--the probability for these being given 
 by a density matrix. For example, at finite temperature, in the real 
 time formalism, the initial conditions are specified at a fixed initial 
 time in the complex t-plane, the probability being given by a Boltzmann weight.
 This is reasonable because in a thermal system operators are of course not 
 ``sandwiched'' between vacuum states.

 We begin again with the full path integral for 
 a massless scalar coupled to an external classical gauge field $A_\mu$.  
 The generating functional is 
 \be
 Z[A_1,A_2] 
 & = &
 \sum_{\psi_f}
 \langle \psi_f | 
 U(t_{\rm fin}, t_{\rm init})\,
 \hat\rho_{\rm init}\,
 U(t_{\rm init}, t_{\rm fin})\,
 | \psi_f \rangle 
 \non
 & = &
 \int_{\cal C} [d\psi]\, \exp\left( i\psi^\dagger D^2 \psi \right)
 \non
 & = &
 \int [d\psi_i]\, 
 [d\psi_f]\,
 {\cal K}[\psi_i]\, 
 Z[A_1]\, Z^*[A_2]
 \label{weights}
\ee
 where the subscript ${\cal C}$ represents the usual Schwinger-Keldysh
 closed time path.  One may consider the two pieces of real time paths
 as representing the two time evolution operators needed to evolve a
 density operator.
 Here, $A_1$ is the background field in the positive time direction, 
 $A_2$ is the field in the negative time direction and
 \be
 Z[A]
 = 
 \int_{\psi_i}^{\psi_f} 
 [d\psi_1]\, 
 \exp\left( i\psi_1^\dagger D^2 \psi_1 \right)
 \ee
 The density operator itself is represented here by the matrix element
 \be
 {\cal K}[\psi_i]
 =
 \langle \psi_i | \hat\rho_{\rm init} | \psi_i \rangle
 \label{eq:calK}
 \ee
 where $\psi_i$ are eigenvectors diagonalizing the Hermitan matrix 
 $\rho_{\rm init}$.  

 To rewrite the above in terms of determinants, we note that $Z[A]$
 can be written as
 \be
 Z[A]^{-1} = \prod_{n}\,\theta_n
 \label{eq:Z_1}
 \ee
 where $\theta_n$ is the eigenvalue of the equation 
 \be
 D^2 \psi_n = \theta_n \, \psi_n 
 \ee
 with the boundary conditions given by $\psi(t_{\rm init})  = \psi_i$
 and $\psi(t_{\rm fin}) = \psi_f$.  

 Taking the logarithm of $Z[A]$ yields 
 \be
 -\log Z[A]
 & = &
 \sum_n  \log \theta_n
 =
 \sum_n  \langle n | \log D^2 | n \rangle 
 \non
 & = &
 \sum_n 
 \int d^4x\, d^4y\,
 \sum_{{\tilde{\lambda}},{\tilde{\lambda}}^\prime}
 \langle n | y,{\tilde{\lambda}}^\prime \rangle \,
 \langle y,{\tilde{\lambda}}^\prime|  \log D^2  | x,{\tilde\lambda}\rangle \,
 \langle x,{\tilde{\lambda}}| n \rangle 
 \non
 & = &
 \sum_n 
 \int d^4x\, d^4y\, 
 \sum_{{\tilde{\lambda}},{\tilde{\lambda}}^\prime}
 \psi_n^\dagger(y,{\tilde{\lambda}}^\prime)\, 
 \langle y,{\tilde{\lambda}}^\prime|  \log D^2  | x,{\tilde{\lambda}} 
 \rangle \,
 \psi_n(x,{\tilde{\lambda}})\nonumber \\
 \label{manybody1}
 \ee
 Here $\tilde\lambda$ is the color label of a particular representation of 
 the algebra\footnote{As in the
 earlier sections, the $\lambda$'s symbolize 
 the eigenvalues of
 the coherent state creation
 and annihilation operators which are generators of arbitrary finite
 dimensional representations of internal symmetries. We can write 
 $\psi_n(y,\lambda^\dagger) = \psi_n(x) + \lambda_a^\dagger \psi_{n,a}(x) + 
 \lambda_a^\dagger\lambda_b^\dagger\psi_{n,ab}(x) \cdots$ as in 
 Ref.~\cite{BalSkag}.}.
 In contrast to the vacuum case, one cannot simply say here that 
 \be 
 \sum_n \psi_n^\dagger(y,{\tilde{\lambda}}^\prime)\,
 \psi_n(x,\tilde{\lambda}) =
 \delta^{(4)}(x-y)\delta_{{\tilde{\lambda}}^\prime\, \tilde{\lambda}} 
 \nonumber
 \ee
 since $\psi_n$ can only span functions with the same boundary 
 conditions.  
 In the vacuum case, one is able to do this because boundary conditions  
 at $t=\pm \infty$ are 
 usually trivially the same for any reasonable observables.
 We now use the identity in Eq.~\pref{trick1}, and re-write
 Eq.~\pref{manybody1} using the same line of reasoning as in section 2.
 (For details of this procedure, see Eqs.~(3.16)--(3.22) in the 
 first paper of Ref.~\cite{DHokerGagne}).
 We obtain 
 \be
 \Gamma[A,\xi] &=& -\log Z[A, \xi]
 \nonumber \\
 &=& 
 \int d^4x\, d^4y\, 
 \sum_{{\tilde{\lambda}},{\tilde{\lambda}}^\prime}
 \xi(x,y,\tilde{\lambda},{\tilde{\lambda}}^\prime)
  \int_0^\infty {dT\over T}{\cal N}\,
  \int_{x,\tilde{\lambda}}^{y,{\tilde{\lambda}}^\prime}\,  {\cal D}z\, 
  \int_{AP} {\cal D}\lambda^\dagger\,{\cal D}\lambda \nonumber \\
&\times& \left({\pi\over T}\right)^n {1\over n}\,\sum_{k=1}^n\, 
 e^{2\pi ik(\lambda^\dagger\lambda +{n\over 2} -1)/n} 
 \exp\left(i\int^T_0 d\tau\, {\cal L}_p (z,\lambda,\lambda^\prime,A)\right)
 \Bigg)
 \, . \nonumber \\
 \label{Npartsc1lp}
 \ee
 We have replaced in the preceding equation
\be
 \sum_n \,\psi_n^\dagger(y,{\tilde{\lambda}}^\prime)\,
 \psi_n(x,\tilde{\lambda}) = 
 \xi(x,y,\tilde{\lambda},{\tilde{\lambda}}^\prime) \, ,
 \ee
 where $\xi$ is a
 functional of both $\psi_i$ and $\psi_f$.
 Also, ${\cal L}_p$ is identical to 
 the point particle Lagrangian in Eq.~\pref{pointLag}.

 Using Eq.~\pref{weights} and Eq.~\pref{Npartsc1lp} above, one obtains the 
 following many body path integral,
 \be
 Z
 = 
 \int [d\xi]\, 
 \exp \left(-G[\xi]\right)\,
 \int_{{\cal C}}\,[dA]\,
 \exp\left(iS_{\rm eff}\right) \, ,
 \label{manybdpath}
 \ee
 where 
 \be 
 S_{\rm eff}[A,\xi] 
 = -\int\, d^4x\, {1\over 4} F_{\mu\nu}^a F^{a,\mu\nu} + \Gamma[A,\xi] \, .
 \label{sceffact}
 \ee
 Here we have made the assumption that the unknown ${\cal K}$ is 
 such that it permits
 Eq.~\pref{manybdpath} with a positive definite $G[\xi]$.  
 We do know from Eq.~\pref{eq:calK} that ${\cal K}$ is real since 
 $\hat\rho_{\rm init}$ is Hermitean.
 What we do not know is whether the Jacobian associated with
 the variable change $(\psi_i, \psi_f) \to \xi$ is real.
 This is not clear since it is not guaranteed that 
 $\xi(x,y,\tilde{\lambda},{\tilde{\lambda}}^\prime)$ is real.
 However we will show shortly that when the saddle point approximation 
 is made to obtain classical world lines, only the real part contributes.
 And in that case, it is likely that the relevant part of $G[\xi]$ is
 positive definite.

 The many body path integral in Eq.~\pref{manybdpath} thus includes an
 average over all possible $\xi$ with the weight $e^{-G[\xi]}$.  Since
 the $\xi$'s here are determined exclusively by solving the
 determinant for the hard modes one arrives at the following physical
 picture.  The above effective action can be thought to represent the
 dynamics of soft modes which are coupled to a background of hard
 modes represented by the point particle world lines. 
 That coupling is weighted by a functional $G[\xi]$. The above
 form of the effective action is reminiscent of the small $x$
 effective action~\cite{smallx1}.  There, by making a separation of
 scales between small $x$ and large $x$, it was shown that $G[\xi]$
 obeyed a non-linear Wilsonian renormalization group
 equation~\cite{smallx2}.  In finite temperature QCD, a similar
 separation of scales occurs at momenta $p\sim T,gT,g^2 T$. Thus
 $G[\xi]\sim G_{\Lambda}[\xi]$, where $\Lambda$ is the scale ($gT\ll
 \Lambda \ll T$) separating hard and soft modes~\cite{Bodeker,BMS}.
 It will be interesting to see if renormalization group
 techniques\footnote{For a discussion of the real time
 renormalization group, see the book by
 Goldenfeld~\cite{Goldenfeld}.}, analogous to those employed in the
 small $x$ effective action, can be used to determine the evolution of
 $G[\xi]$ as a function of the cut-off at finite T~\cite{Boyanovsky,Litim}.

 We should comment on the real time contour ${\cal C}$ in the path
 integral over the effective action in Eq.~\pref{manybdpath}. In the
 real time formalism of quantum field theory, for thermal initial
 conditions, the path consists of a piece where the field theory is
 evolved in imaginary time from $t=0$ to $t=-i\beta$ and two time
 slices from $-\infty$ to $+\infty$ and back from $\infty-i\beta$ to
 $-\infty -i\beta$. Propagators have a $2\times 2$ matrix structure,
 with the off--diagonal terms corresponding to a flux of on--shell,
 thermal particles. The corresponding problem of many body propagators
 in the world line formalism has been addressed by
 Mathur~\cite{Mathur}. He considers a more general initial state than
 a thermal one, with the only restriction on the initial density
 matrix being the constraint that it can be represented as $\rho =
 e^{B[\phi]}$, where $B$ is quadratic in the field operator
 $\phi$. The thermal distribution is a particular instance of such
 `exponential of quadratic' density matrices, which have the property
 that correlators computed with these matrices, satisfy Wick's
 theorem.

 In the first quantized language of world lines, one obtains the following 
 picture~\cite{Mathur}. Since the world line particle moves only in one 
 direction in proper time, the paths backwards in time correspond to 
 switching the sign of the einbein discussed in section 2. Specifying the 
 amplitude $\xi$ to a reverse orientation of proper time is equivalent to 
 specifying an `exponential of quadratic' density matrix. Replacing $\xi$ by a  $\delta$--function, as discussed previously, would let us recover the 
 Feynman propagator. For the particular case of classical transport theory 
 in hot QCD to be discussed in the next section, we will not need to specify 
 the real time contour, but it will be relevant for future detailed 
 computations.   

 To summarize, in this section, a function $\xi$ is
 obtained, and a path integral introduced over all $\xi$ with an
 unspecified weight $G[\xi]$. Effectively, what one has done is to introduce 
 a density matrix $\tilde{\rho}$ corresponding to the hard modes in the 
 system. The density matrix of course is required to satisfy the condition 
 \be
 {\rm Tr}\left(\tilde{\rho}\right) 
 = \int\,[d\xi]\, \exp\left(-G[\xi]\right) 
 = 1
 \, .
 \ee

 \section{From the QCD one loop effective action to classical
 transport theory}

 In the previous section, we extended the world line formalism
 discussed in section 2 to a many body context.  Our objective in this
 section is to derive from the one loop effective action 
 in many body QCD
 under suitable approximations, the 
 classical transport theory of soft classical modes coupled to the world 
 lines of the hard modes.
 In particular, we will apply the world line formalism to gauge theories 
 at the very high temperatures where weak coupling techniques are applicable.
 We will see that we recover the coupled set of transport equations that 
 were written down by Litim and Manuel~\cite{LitimManuel}.
 It was shown 
 by these authors that their results, in turn, agreed with the earlier work, 
 in an apparently very different approach by B\"{o}deker~\cite{Bodeker}.

 In hot QCD, the leading high $T$ contribution to the one loop effective
 action for gluons is the same as the action for the example we
 discussed in previous sections--namely, that of a complex scalar
 field in the adjoint representation. We see this as
 follows~\cite{EHZ}.  Working in the background field gauge, and
 integrating out the fluctuations to quadratic order in the hard
 fields, one obtains
 \be
 S_{\rm eff} = S_{\rm soft} + {\rm Tr}\left(\log(-D^2)\right) - {1\over 2} {\rm 
 Tr} \left(\log\left[-D^2 g_{\mu\nu} + 2 F_{\mu\nu}\right]\right) \, ,
\label{turnips} 
\ee
 where $D_\mu$ is the usual covariant derivative for the background
 field.  The above equation will also have in general a gauge fixing
 term-with a gauge fixing parameter $\alpha$. For simplicity, we have
 eliminated this term by setting Feynman gauge, $\alpha=1$.  In 
 Ref.~\cite{EHZ}, it was shown that the $\alpha$ dependence
 is suppressed by powers of $1/T$.
 
 One can expand the above equation, writing
 \be
 S_{\rm eff} &=& S_{\rm soft} - {\rm Tr}\left(\log(-D^2)\right) - {\rm 
 Tr} \left(\log\left[{2 F_{\mu\nu}\over D^2}\right]\right)\nonumber \\
 &+&
 {1\over 2} {\rm Tr} \left( {1\over D^2}\,2 F^{\mu\rho}{1\over D^2}\, 
 2F^{\rho\nu}\right)+ \cdots \nonumber \\
 &\sim& S_{\rm soft} + \Gamma[A] \, .
 \ee
 By power counting arguments~\cite{BraatenPisarski}, the terms
 containing $F^{\mu\nu}$ are suppressed by powers of $T$ at high
 temperature. Hence, $\Gamma[A]$ for hot QCD is the same as the
 $\Gamma$ we derived in the previous section.

 We now wish to derive the classical transport theory that follows from 
 the above effective action. If we insert a functional derivative 
 $\delta/\delta A$ in  Eq.~\pref{manybdpath}, we have the identity
\be
  \int \, [d\xi]\, 
 \exp \left(-G[\xi]\right)\,
 \int_{{\cal C}}\,[dA]\,
 {\delta\over \delta A^\nu}\,\exp\left(iS_{\rm eff}\right)=0 \, .
\label{opident}
 \ee
This is of course equivalent to the operator equations of motion
\be
 \langle D_\mu F^{\mu\nu} - J^\nu \rangle \equiv {\rm Tr}\left(\tilde{\rho}\,
 \left(D_\mu F^{\mu\nu}-J^\nu\right)\right)=0 \, .
\label{ymexptvalue}
 \ee
 From Eq.~\pref{opident} and Eq.~\pref{Npartsc1lp}, one obtains 
\be
 \langle J^{\nu,\alpha}(x)\rangle &=& \int \, [d\xi]\,
e^{ -G[\xi]} \,\int_{{\cal C}}[dA]\,e^{iS_{\rm eff} } \,   
\int dr\, dy\, \sum_{\tilde{\lambda},{\tilde{\lambda}'}} 
\Bigg(
 \xi(r,y,\tilde{\lambda},{\tilde{\lambda}}^\prime) \nonumber \\
  &\times& \int_0^\infty {dT\over T}\, {\cal N}\,
  \int_{r,\tilde{\lambda}}^{y,{\tilde{\lambda}}^\prime}\, 
{\cal D}z\, 
  \int_{AP} {\cal D}\lambda^\dagger\,{\cal D}\lambda\,
 \left({\pi\over T}\right)^n \,{1\over n}\,
\sum_{k=1}^n\, 
 e^{{2\pi i k\over n}\cdot (\lambda^\dagger\lambda +{n\over 2} -1)} \nonumber \\
&\times&  \int_0^T \,d\tau \, g\,\lambda^\dagger\,T^{\beta}\,\lambda\, {\dot{z}}_\sigma\,
 {\delta A^{\sigma,\beta}(z(\tau))\over \delta A^{\nu,\alpha}(x)} \, \,
 e^{i\int_0^T\,d\tau' {\cal L}_p (z,\lambda,
\lambda^\dagger,A)} \Bigg) \, . \nonumber \\
\label{therealcurrent}
\ee

Let's now take the saddle point of the Quantum Mechanical path 
integral in the expectation value of the 
current\footnote{The sum $\sum_{k=1}^n e^{{2\pi i k\over n}\cdot 
(\lambda^\dagger\lambda +n/2 -1)}$ 
in the path integral in Eq.~\pref{therealcurrent} 
is analogous to the well known GSO projection 
in string theory. As discussed in section 3 (after Eq.~\pref{DComm}), this 
projection restricts the world line paths to lie in a particular irreducible 
representation of the group. We will assume implicitly hereafter that this 
restriction is obeyed by the classical path.}.
At present, this step can only be justified {\it a posteriori}. However, it
can be anticipated. One expects the world lines of the hard $p\sim T$ modes 
to obey classical trajectories {\it a la Wong} 
since their coupling to the soft modes is 
weak at very high temperatures. (This also happens to be the case for the 
coupling between small $x$ and large $x$ modes in the small $x$ effective 
action.)
This information should of course be 
obtained self-consistently from the effective action. We hope to address the 
validity of this classical approximation at a later date. 

Taking the saddle point of the Quantum Mechanical path integral, we obtain 
\be 
\langle J^{\mu,\alpha}(x)\rangle  = \int \, [d\xi]\,\exp
\left(-G[\xi]\right)\, \int_{\cal C}\,
[dA]\,\exp\left(iS_{\rm eff}\right)\, J_{\xi, A}^{\mu,\alpha}(x) \, ,
\label{classicalJ0}
\ee
where
\be
 J_{\xi,A}^{\mu,\alpha}(x) &=&  
\int dr\, dy\, \sum_{\tilde{\lambda},{\tilde{\lambda}'}}
 \xi(r,y,\tilde{\lambda},{\tilde{\lambda}}^\prime)\,
 {\cal N}^\prime\,
 \int_0^\infty {dT\over T}
 \left({\pi\over T}\right)^n\,{1\over n}  
 \non & & \qquad {} \times
 \exp\left(i\int_0^T\,d\tau^\prime {\cal L}_p (z_c,\lambda_c,
 \lambda_c^\dagger,A)\right) 
 \non & & \qquad {} \times
 \int_0^T \, d\tau \, g\,
 \lambda_c^\dagger\,T^{\alpha}\,\lambda_c\, {\dot{z}}_c^\nu\,
 \delta^{(4)}(x-z_c(\tau))\,
 \non
 \label{classicalJ1}
 \ee
 In Eq.(40) the effective action is now defined to be 
 $S_{\rm eff} = S_{\rm soft} + \Gamma_{\rm cl}$
 where 
 \be
 \Gamma_{\rm cl}
 & = &
 \int dr\, dy\, \sum_{\tilde{\lambda},{\tilde{\lambda}'}}
 \xi(r,y,\tilde{\lambda},{\tilde{\lambda}}^\prime)\,
 {\cal N}^\prime\,
 \int_0^\infty {dT\over T}
 \left({\pi\over T}\right)^n  
 \exp\left(i\int_0^T\,d\tau^\prime {\cal L}_p (z_c,\lambda_c,
 \lambda_c^\dagger,A)\right) 
 \non
 \ee
The subscripts $c$ on the variables correspond to the classical paths which 
have the endpoints $(r,\tilde{\lambda})$ and $(y,{\tilde{\lambda}}^\prime)$. 
Note that for the classical path 
the $\tau$ integral is really independent of $T$ because
rescaling $\tau \to T\tau$ does not change anything. (Recall from section 3 
that the classical point particle Lagrangian is reparametrization invariant.)
Hence, extending the range of integration, 
$0<\tau < T\,\longrightarrow \, 0<\tau < \infty$ should not modify the 
result.
Performing this operation results in 
\be
J_{\xi,A}^{\mu,\alpha}(x)& & = \nonumber \\
 g& &\!\!\!\!\! 
\int dr\, dy\, \sum_{\tilde{\lambda},{\tilde{\lambda}'}}
\Bigg(
 \xi(r,y,\tilde{\lambda},{\tilde{\lambda}}^\prime)
 \int_0^\infty \, d\tau \,
 \Bigg\{ Q^\alpha (\tau,r,y,\tilde{\lambda},{\tilde{\lambda}}^\prime)
\, {\dot{z}}_c^\mu (\tau,r,y,\tilde{\lambda},
{\tilde{\lambda}}^\prime)\nonumber \\
\times & &\!\!\!\!\! \delta^{(4)}(x-z(\tau,r,y,\tilde{\lambda},{\tilde{\lambda}}^\prime))
\,\Bigg\} \Bigg)\, .
\label{classicalJ2}
\ee
We
have here used Eq.~\pref{Qdef} to introduce the color charge $Q^\alpha$.
Also, we have redefined $\xi$ to absorb the infinite normalization
constant coming from $\cal N^\prime$ and the $T$ integral. 
A discussion of the renormalization of infinities in the first quantized 
formalism is outside the scope of this paper. We refer interested readers 
to the discussion in Ref.~\cite{PerBo}.

Now introduce the dummy integrals 
\be
\int\, dQ\,\delta^{(n)}(Q-Q(\tau))
=1 \ \ \ \ {\rm and} \ \  \int\, d^4p\, \delta^{(4)}(p-p(\tau))=1 \, . 
\ee
The color charge measure 
imposes the constraint (see footnote 7) that the Casimir invariants 
for the group are conserved. For SU(3)~\cite{Kellyetal}, 
\be
dQ = d^{(8)}Q\,\delta(Q_a Q^a -q_2)\, \delta(d_{abc}Q^a Q^b Q^c -q_3) \, ,
\ee
where $q_2$ and $q_3$ fix the values of the Casimir invariants, and $d_{abc}$ 
are the symmetric constants of the group. The color charges spanning the 
phase space are dependent variables but, as argued by Kelly et al., they 
can be formally related to a set of independent phase-space Darboux 
variables~\cite{Alexeev}. We will not further 
address the issue of the color measure in this paper.

Substituting the dummy integrals for $p$ and $Q$, making use of Wong's 
equations, and switching the order of integration, the current can now be 
re-written as 
\be
J_{\xi,A}^{\mu,\alpha}(x) =  g\,\int\, d^4 p\,dQ\, p^\mu \,Q^\alpha \, 
f_{\xi,A}(x,p,Q) \, .
\label{classicalJ3}
\ee 
We have introduced a function $f_{\xi,A}(x,p,Q)$ which is defined by the 
relation
\be
f_{\xi,A}(x,p,Q) &=& \int_0^\infty \, d\tau\, 
\int dr\, dy\, \sum_{\tilde{\lambda},{\tilde{\lambda}'}}
\Bigg(\xi(r,y,\tilde{\lambda},{\tilde{\lambda}}^\prime)\,\delta^{(4)}
(x-z(\tau,r,y,\tilde{\lambda},{\tilde{\lambda}}^\prime))\nonumber \\
&\times&
\delta^{(n)}(Q-Q(\tau,r,y,\tilde{\lambda},{\tilde{\lambda}}^\prime))\,
\delta^{(4)}
(p-p(\tau,r,y,\tilde{\lambda},{\tilde{\lambda}}^\prime))\,\Bigg) \, .
\label{microdist}
\ee

Consider now what $f_{\xi,A}$ represents.
First note that written in this way, it is clear that only the real part of 
$\xi$ contributes to the color current.  This can be easily shown by
noticing that
$\xi(r, y, \tilde{\lambda}, \tilde{\lambda}')
=
\xi^*(y, r, \tilde{\lambda}', \tilde{\lambda})$ and renaming
$r \leftrightarrow y$, $\lambda \leftrightarrow \lambda'$.
Classical paths $z$ and $Q$ are not affected by this exchange 
because given the boundary conditions, these are fixed. 
The meaning of
$\xi(r, y, \tilde{\lambda}, \tilde{\lambda}')$ is that of the probability 
density of a particular path that begins with  
$(r, \tilde\lambda)$ and ends with $(y, \tilde\lambda')$.  
However, since we are
dealing with classical paths here, one can equally well specify a
trajectory by the position and the velocity and the value of the charge
at any fixed time\footnote{If we do not take the classical saddle point, 
both boundary conditions will matter.}. 
(Note that since Eq.~\pref{eq:Qdot} is a first order
differential equation, one needs only one boundary condition.)
In particular, we can do it at $\tau = 0$.  If we perform the trace 
explicitly, there will be Jacobian corresponding to the affine 
transformation. For a discussion of this point, we refer the reader to a 
paper by Brandt, Frenkel, and Taylor~\cite{BFT}--in particular Eq.~(2.14) 
therein.

Returning to Eq.~\pref{microdist}, one can show that $f_{\xi,A}$
satisfies the 
equation\footnote{For an analogous explicit derivation, see Appendix A of 
Ref.~\cite{Kellyetal}}
\be
p^\mu \left( {\partial \over \partial x^\mu} -2g{\rm Tr} (QF^{\mu\nu}) 
{\partial \over \partial p^\nu} + 2g {\rm Tr}([A_\mu,Q]{\partial\over 
\partial Q})\right) f_{\xi,A}(x,p,Q) =0 \, .
\label{eq:vlasov}
\ee
Although it appears as such, the above equation is not quite the
collisionless Vlasov equation, since $f_{\xi,A}$ here is
a microscopic quantity and 
not a single particle distribution in the statistical mechanics sense.
 
To obtain an equation for the single particle distributions, 
averages both over $\xi$ and $A$ 
must be performed.  
Keeping in mind that the expressions here are functionals of $A$ and $\xi$, 
we have
\be
0
& = &
p^\mu
\left( 
{\partial \over \partial x^\mu} 
\left\langle f_{\xi,A}(x,p,Q) \right\rangle 
-
2g{\rm Tr} \left\langle (Q F^{\mu\nu} ) 
{\partial \over \partial p^\nu} f_{\xi,A}(x,p,Q) \right\rangle 
\right.
\non
& & {} \qquad
- 
\left.
2g {\rm Tr}\left\langle 
([Q, A_\mu]{\partial\over \partial Q})f_{\xi,A}(x,p,Q)
\right\rangle 
\right) 
\label{eq:master1}
\ee
Here the bracket $\langle \cdots \rangle$ indicates the functional 
averages over $A$ and $\xi$ with the weights specified in 
Eq.~\pref{classicalJ0}.
The corresponding equation for $A$ is
\be
\langle D_\mu F^{\mu\nu} \rangle = \langle J^\nu \rangle
\label{eq:master2}
\ee
where $\langle J\rangle$ is expressed in terms of $f_{\xi,A}$ via 
Eqns.~\pref{classicalJ0} and \pref{classicalJ3}.
Note that the left hand side of this equation consists of 2 and 3 point
functions of $A$.

The equations \pref{eq:master1} and \pref{eq:master2}
are our main results in their most general form. They are not the 
Vlasov equations as noted
by Litim and Manuel\cite{LitimManuel}. This is simply because these equations
know about 2 and 3 point correlations, whereas in the Vlasov equation only the
average values appear.  Ignoring these correlations, one would get the
Vlasov equation.

To consider fluctuations, following Litim and Manuel, we shall now define  
\be
f_{\xi,A} = \langle f \rangle + \delta f
\label{eq:barf}
\\
\inline{and}
A = \langle A\rangle + a
\label{eq:barA}
\ee
where $\langle f\rangle$ and $\langle A\rangle$ 
denote the functional average of $f_{\xi,A}$ and $A$
in the path integral for the effective action (see Eq.~\pref{classicalJ0}). 
Note that by definition
$\langle \delta f\rangle$ and $\langle a \rangle$ both vanish.
Furthermore, one has
\be
F^{\mu\nu} = \bar{F}^{\mu\nu} + f^{\mu\nu} \, ,
\ee
where 
\be
\bar{F}^{\mu\nu}
& = &
\partial^\mu \langle A \rangle^{\nu,a} 
-\partial^\nu \langle A \rangle^{\mu,a} 
+g\,f^{abc} \, \langle A \rangle^{\mu,b}\, \langle A \rangle^{\nu,c} 
\non
f^{\mu\nu,a} 
& = & 
({\bar D}^\mu a^{\nu})^a 
-
({\bar D}^\nu a^{\mu})^a + g\,f^{abc} 
\,a^{\mu,b}\,a^{\nu,c} 
\ee
with the covariant derivative ${\bar D}_\mu$ defined to be $\partial_\mu - 
ig\,\langle A_\mu\rangle$.

Substituting Eqs.~\pref{eq:barf} and \pref{eq:barA} into
Eqs.~\pref{eq:master1} we obtain 
\be
p^\mu\left( 
{\bar D}_\mu \langle f_{\xi,A}(x,p,Q)\rangle -
2g\, {\rm Tr}(Q \bar{F}^{\mu\nu} ) 
{\partial \over \partial p^\nu} \langle f_{\xi,A}(x,p,Q)\rangle 
\right) = \langle \eta\rangle + \langle \zeta\rangle \, .
\label{eq:langevin1}
\ee
In the preceding equation, $\langle \eta \rangle$ and $\langle \zeta\rangle$ 
are given by 
\be
\langle \eta \rangle &=& g\,p^\mu\,Q^a\, \langle f_{\mu\nu}^a\,\partial_p^\nu 
\delta f\rangle \nonumber \\
\langle \zeta \rangle &=& g\,Q^a\,f^{abc}\,p^\mu\,\langle a_\mu^b a_\nu^c\rangle \partial_p^\nu \langle f\rangle + g\,f^{abc}\,Q^b\,p^\mu\,\langle a_\mu^c\,
\partial_{Q^a}\delta f\rangle \, ,
\label{eq:lang1RHS}
\ee
and 
\be
{\bar D}_\mu \langle f_{\xi,A}(x,p,Q)\rangle = \left(\partial_\mu -
g\,f^{abc}\,Q^c\,A_{\mu}^b\,\partial_{Q^a}\right)\,
\langle f_{\xi,A}(x,p,Q)\rangle \, .
\ee

Performing the same substitution in Eq.~\pref{eq:master2}, we find 
\be
{\bar D}_\mu \bar{F}^{\mu\nu,a} +\langle J_{\rm fluct}^{\nu,a}\rangle 
= \langle J^{\nu,a}\rangle \, ,
\label{eq:langevin2}
\ee
with $\langle J_{\rm fluct}^{\nu,a}\rangle$ defined by the relation 
\be
\langle J^{\nu,a}\rangle = g\,f^{abc}\,\left\langle a_\mu^b\left(
\left( ({\bar D}^\mu a^\nu)^{c} - ({\bar D}^\nu a^\mu)^{c} \right)
+g\,f^{cde}\,a_d^\mu a_e^\nu
\right) + ({\bar D}_\mu a^\mu a^\nu)^{bc}\right\rangle \, .
\label{eq:lang2LHS}
\ee

Proceeding from the preceding set of equations, Litim and Manuel have
performed an extensive analysis, in weak coupling, of these equations
and their solutions. For instance, neglecting the fluctuation terms introduced 
above, and expanding single particle distribution about the equilibrium 
Bose or Fermi  distribution $\langle f_{\rm equil}\rangle$,
\be
\langle f(x,p,Q)\rangle = \langle f_{\rm equil}^{(0)}(p_0)\rangle + g\langle 
f^{(1)}(x,p,Q)\rangle + \cdots \, ,
\ee
one can reconstruct the hard thermal loop effective action, as previously 
shown by Kelly et al. Keeping the fluctuation terms, and similarly expanding 
$\langle \eta\rangle$, $\langle \zeta\rangle$ and \
$\langle J_{\rm fluct}\rangle$, 
one obtains the Boltzmann--Langevin equation containing both noise and 
collision terms that was first derived by B\"{o}deker~\cite{Bodeker}.

We will not perform such an analysis in this section, and wish to
direct interested readers to the papers mentioned. The objective of
this section was to demonstrate that the starting point for such
analyses can be obtained, after clear approximations, from the many
body world line formalism developed in this paper. It is hoped that
this formalism will allow us to understand better the validity and the
limitations of the various approximations employed, and how one may go
beyond these approximations. 

In particular, it will be useful to
better understand the following: a) when does the saddle point
approximation for the quantum mechanical path integral break down?
Presumably, in this case, the quantum problem can be reformulated in
terms of Wigner functions.  b) It appears that in this formalism one
may be able to distinguish between thermal and quantum
fluctuations. How is this achieved in practice? c) The collision-less Vlasov 
equations can be exponentiated to obtain the hard thermal loop effective 
action. The addition of damping and noise terms, as B\"{o}deker has 
emphasized, implies that 
we are going beyond hard thermal loops. Does consistency then require that 
one keep the sub--leading terms in Eq.~\pref{turnips}? d) What is the form
of the functional $G[\xi]$? Is a renormalization group treatment,
wherein $G$ encapsulates information about non--trivial correlations
feasible?  
We hope to address some of these issues, in the finite
temperature context, in a future work~\cite{JSRJ}.

\section{Discussion}
 
The world line formalism in quantum field theory, going as far back as
the early works of Feynman and Schwinger, provides a direct
representation of quantum field theory in terms of world lines of
point particles. Recently, it has become popular as a powerful method
to compute Feynman diagrams in theories with internal symmetries such
as spin and color. These internal symmetries are treated using
classical Grassmanian variables. 

In the many body problem, the world line formalism leads to
an intuitive picture consistent with our ideas about 
classical kinetic theory--that of phase space averages over the world lines 
of classical particles. We have shown in this paper that classical transport 
theory can be obtained form the many body world line formalism. As discussed 
by several authors, this classical transport theory, in the Vlasov limit 
sums the leading class of Feynman diagrams for external momenta $p\sim gT$--
called ``hard thermal loops''. 
The inclusion of noise and damping terms to give the Boltzmann--Langevin 
equation, goes beyond hard thermal loops and allows one to address physics 
at the softer scale $p\sim g^2 T$. The connections of the imaginary time 
world line formalism to classical transport theory is fleshed out for the 
specific example of a $\phi^4$ theory in appendix A.

The potential advantage of the formalism developed in this paper is that 
one has a formalism at the level of the effective action. It may therefore 
facilitate computations of effects beyond those of classical transport theory. 
Its form suggests that it is amenable to a ``dynamical renormalization group'' 
treatment, as has been discussed previously by Boyanovsky, de Vega, and 
collaborators~\cite{Boyanovsky}. (See also very recent work, the many body 
extension of which may lead to a picture similar to ours~\cite{Morris}.) 
Similar ideas have been used previously
in small $x$ physics to develop a Wilsonian renormalization group for 
wee partons~\cite{smallx1,smallx2}. Indeed, one can draw a formal analogy 
between the $P$'s and $Q$'s in classical transport theory, and the classical 
sources in the small $x$ effective action. Despite the formal similarity, 
one should caution that 
the physics in the two cases is quite different. Unlike classical 
transport theory, small $x$ physics is dominated by light cone singularities. 
One consequence is that the light cone sources are static, not dynamic like 
the $P$'s and the $Q$'s. It is therefore not obvious that the renormalization 
group ideas of small $x$ physics can be transported `lock, stock and barrel' 
over to the physics of hot QCD. 

Nevertheless, there are connections that are worth exploring further. A. H. 
Mueller has shown recently that the classical effective action, when applied 
to nuclear collisions~\cite{AlexRaj}, leads to the Boltzmann 
equation~\cite{AHMueller}. Also, recently Elmfors, Hansson, and Zahed have 
shown that the hard thermal loop free energy could be related directly, 
via the relativistic virial expansion, to the S--matrix elements of high 
energy scattering~\cite{EHZ}.

We should point out that we have glossed over many subtleties in the
world line formalism. In particular, how the closed time path of
quantum field theory manifests itself in the first quantized formalism
needs to be better understood. Whether one can separate statistical
and quantum fluctuations, is another topic that should be studied
further. Another issue that needs to be addressed is whether one can
go beyond the one loop approximation in the world line formalism. This
has been achieved for the vacuum case by Schubert and
collaborators~\cite{HellSchool}, and one should be able to extend it
in a similar fashion to the many body context. Finally, there are a large 
number of physical problems that should be hopefully simpler to compute in 
the many body world line formalism.

\section*{Acknowledgements}
We would like to thank Hans Hansson, Edmond Iancu, Cristina Manuel
and Rob Pisarski for useful discussions, and for their comments on the
manuscript.  We would also like to acknowledge useful discussions with
Paulo Bedaque, Dan Boyanovsky, Gerald Dunne, Alex Krasnitz, Daniel
Litim, Dam Son, and Ismael Zahed. Finally, we would like to thank Rob
Pisarski for introducing us to the problem, and for insisting that we
mind our P's and Q's.  S. J.~and J.J.-M. were supported by the
Director, Office of Energy Research, Office of High Energy and Nuclear
Physics, Division of Nuclear Physics, and by the Office of Basic
Energy Sciences, Division of Nuclear Sciences, of the U.S. Department
of Energy under Contract No. DE-AC03-76SF00098. R. V. is supported by
DOE Nuclear Theory at BNL. J.J-M and J. W. thank the Nuclear Theory group at
BNL for their hospitality.

\appendix

\section{The effective action in the imaginary time world line formalism}

In this appendix we derive the leading term in the finite
temperature effective action for a $\phi^4$ theory,
using directly the one loop world line formalism at imaginary time.

Starting from the full imaginary time action,
\be 
S = \int d^4 \! x 
\left [ -\frac{1}{2} \phi \partial^2 \phi + \lambda \phi^4 \right ] 
\ , \label{phistart}
\ee
where $\int d^4 \! x = \int_0^{\beta} d\! \tau \int d^3 \! x$ 
and $\beta$ the inverse temperature $T$, 
we split the field $\phi$ into a ``soft'' and ``hard'' part, 
$\phi = \phi_s + \phi_h$.
In the one loop approximation, the action Eq.~\pref{phistart} becomes,
\be
S_{{\rm 1-loop}} = \int d^4 \! x 
\left [ -\frac{1}{2} \phi_s \partial^2 \phi_s + \lambda \phi^4_s 
+ \frac{1}{2} \phi_h \left (-\partial^2 +12\lambda
\phi^2_s \right ) \phi_h \right ] \ .
\ee
Integrating out the hard part, we obtain the effective action for the 
soft field,
\begin{equation}
S_{{\rm eff}} =  \int d^4 \! x 
\left [ -\frac{1}{2} \phi_s \partial^2 \phi_s + \lambda \phi^4_s \right ] 
+\frac{1}{2}
{\rm LogDet} \left ( -\partial^2 +12\lambda \phi^2_s \right) 
= S_{{\rm eff}}^{(1)} + S_{{\rm eff}}^{(2)} \, . \nonumber 
\\ \label{effaction} 
\end{equation}

\vspace{2mm}
To evaluate the remaining determinant, we use the world line approach 
at finite temperature (and imaginary time), developed in 
\cite{mckeonrebhan1}, and write the second term in 
Eq.~\pref{effaction} as
\be
S_{{\rm eff}}^{(2)} &=& -\frac{1}{2} 
\sum_{n=-\infty}^{\infty} \int_0^{\infty} \frac{dt}{t} 
\int {\cal D} \! p \,
e^{-\int_0^{t} dt_1 p^2} \, \times \nonumber \\ &&
\int_{P_T} {\cal D} \! x \, 
\exp \left[ -\int_{0}^{t} dt_1 \left \{ \frac{\dot{x}^2}{4} 
+12\lambda \phi^2_s(x(t_1)) \right \} \right ] \ . \label{worldlinestart}
\ee   
The boundary condition $P_T$ in the path integral is given by
\be
x_{\mu} (t) = x_{\mu} (0) + n \beta \delta_{\mu 4} \ , \label{theboundcond}
\ee
and for simplicity we have put the einbein $\varepsilon = 2$.

The term corresponding to the blackbody radiation is easily obtained 
by taking $\lambda \rightarrow 0$ and making the substitution
\be
x_{\mu}(t_1) = u_{\mu}(t_1) + z_{\mu} + \frac{n\beta t_1}{t}\delta_{\mu 4} \ , 
\label{coordsubst}
\ee
where now $u_{\mu}(t) = u_{\mu}(0) =0$ and $z_{\mu}$ is $t_1$--independent. We obtain,
\be
S_{{\rm eff}}^{(2)}(\lambda =0) &=& 
-\frac{1}{2} \int d^4 \! z \sum_{n=-\infty}^{\infty} \int_0^{\infty} \frac{dt}{t} 
e^{-n^2 \beta^2 / 4t}
\int {\cal D} \! p \, e^{-\int_0^{t} dt_1 p^2} \times 
\nonumber \\ && \int {\cal D} \! u \, e^{-\int_{0}^{t} dt_1 
\left ( \dot{u}^2 /4 \right )} 
=  \frac{-1}{32\pi^2} \int d^4 \! z \sum_{n=-\infty}^{\infty} 
\int_0^{\infty} \frac{dt}{t} e^{-n^2 \beta^2 / 4t} \ , 
\nonumber \\ \label{freepart}
\ee
where we have used the explicit result for a gaussian path integral. 
The finite temperature part of Eq.~\pref{freepart} is given by
\be
S_{{\rm eff}}^{(2)}(\lambda =0) &=& -\frac{1}{16\pi^2} 
\int d^4 \! z \sum_{n=1}^{\infty} \int_0^{\infty} dv \, 
v e^{-n^2 \beta^2 v /4} 
\nonumber \\ &=& -\frac{T^4}{\pi^2} 
\int d^4 \! z \sum_{n=1}^{\infty} \frac{1}{n^4} = 
-\frac{\pi^2 T^4}{90} \int d^4 \! z \ ,  
\label{finalblack}
\ee
which is just the result for a noninteracting gas of scalar particles. 
The corresponding result for a
free gluon gas is a straightforward generalization of the above procedure,
using Eq.~\pref{DHoker} to represent the internal degrees of freedom.

To obtain the first nontrivial result, we calculate the corresponding current,
\be
j(y) &=& \frac{\delta S_{{\rm eff}}^{(2)}}{\delta \phi_s (y)} 
= 12\lambda \phi_s (y) \sum_{n=-\infty}^{\infty} 
\int_0^{\infty} \frac{dt}{t} \int {\cal D} \! p \, 
e^{-\int_0^{t} dt_1 p^2} \int_{P_T} {\cal D} \! x \times \nonumber \\ && \exp \! 
\left[ -\int_{0}^{t} dt_1 \left \{ \frac{\dot{x}^2}{4} 
+12\lambda \phi_s^2 (x(t_1))
\right \} \right ] \left ( \int_0^t dt_2 \delta^4 (y-x(t_2)) \right ) , 
\nonumber \\ \label{currentstart}
\ee
where the delta function is periodic in the imaginary time, with period $\beta$.
Note that this expression is nothing but the average value of the single 
particle current, in the background field 
$\phi_s$. Thus, it corresponds to the imaginary time formulation of 
Eq. \pref{therealcurrent}, 
although for a $\phi^4$ theory in the present case.
 
To lowest order, we may neglect the $\lambda$-dependence 
in the world line action. Using also the substitution
in Eq. \pref{coordsubst}, the current becomes,
\be
j(y) &\rightarrow & j^{(1)} (y) = 12\lambda \phi_s (y) \sum_{n=-\infty}^{\infty} 
\int_0^{\infty} \frac{dt}{t} e^{-n^2 \beta^2 / 4t}
\int {\cal D} \! p e^{-\int_0^{t} dt_1 p^2} \times \nonumber \\ &&
\int {\cal D} \! u e^{-\int_{0}^{t} dt_1 
\left ( \dot{u}^2/4 \right ) } \int
d^4 \! z \left ( \int_0^t dt_2 \delta^4 (y-u(t_2) -z -n\beta t_2 /t) \right ) 
\ . \nonumber \\ \label{current2}
\ee
Now, since
\be
\int d^4 \! z \left ( \int_0^t dt_2 
\delta^4 (y-u(t_2) -z -n\beta t_2 /t) \right ) =
\int_{0}^{t} dt_2 = t \ ,
\ee
the current in Eq. \pref{current2} becomes
\be
j^{(1)}(y) &=& \frac{3\lambda}{4\pi^2} \phi_s (y) 
\sum_{n=-\infty}^{\infty} \int_0^{\infty} \frac{dt}{t^2} 
e^{-n^2 \beta^2 / 4t} \ ,
\ee
and the finite temperature part is 
\be
j^{(1)}(y) &=& \frac{6\lambda T^2}{\pi^2} \phi_s (y) 
\sum_{n=1}^{\infty} \frac{1}{n^2} = \lambda T^2 \phi_s (y) \ .
\label{finalcurrent}
\ee

>From the first part of Eq. \pref{effaction}, 
the blackbody radiation in Eq. \pref{finalblack} 
and the current in Eq. \pref{finalcurrent}, we finally
get the lowest order form of the effective action for the soft field, 
\be
S_{{\rm eff}} = \int d^4 \! z 
\left [ -\frac{1}{2} \phi_s \partial^2 \phi_s + \lambda \phi^4_s - 
\frac{\pi^2 T^4}{90} + \left ( \frac{\lambda T^2}{2} \right ) \phi_s^2 
\right ] \ . \label{finaleff}
\ee

The leading term in the effective action corresponds to nothing 
but the thermal mass for the soft field,
as depicted in Fig. 1a. In principle the thermal mass 
also depends on the infrared cutoff, i.e. the cutoff that characterizes the 
separation between the hard and soft fields, as discussed in \cite{BMS}. However, this term is subleading
compared to $T^2$ and has been neglected in the present treatment.
 
Within the one loop approximation, higher powers of the soft field come 
from diagrams of the form shown in Fig. 1b, and in general there are of course also corrections
from higher loop diagrams.   

\begin{center} \begin{picture}(320,100)(0,0)
\Line(0,30)(80,30)
\Vertex(40,30){3}
\DashCArc(40,50)(20,0,360){4}
\Text(40,20)[c]{a:}
\DashCArc(180,50)(20,0,360){4}
\Line(120,70)(160,50)
\Line(120,30)(160,50)
\Vertex(160,50){3}
\Line(240,70)(200,50)
\Line(240,30)(200,50)
\Vertex(200,50){3}
\Vertex(154,65){1}
\Vertex(180,80){1}
\Vertex(159,71){1}
\Vertex(165,76){1}
\Vertex(172,79){1}
\Vertex(188,79){1}
\Vertex(195,76){1}
\Vertex(201,71){1}
\Vertex(206,65){1}
\Text(180,20)[c]{b:}
\Line(260,60)(280,60)
\Text(285,62)[l]{= soft field}
\DashLine(260,40)(280,40){4}
\Text(285,42)[l]{= hard field}
\end{picture}  
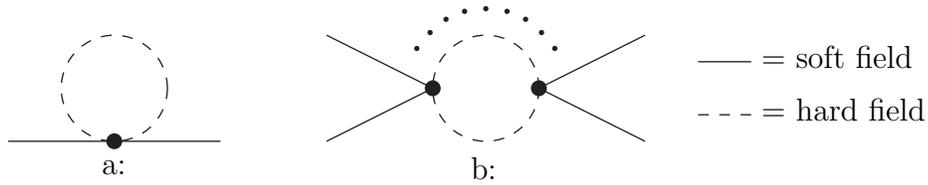
\\ {\sl {\rm Fig. 1.} {\rm a:}Thermal mass for the soft field. 
\\ \hspace{3.1cm} {\rm b:} Higher corrections
to the effective action}
\end{center}

Let us now briefly compare the above world line formulation with the
classical kinetic theory, as studied in \cite{phi4boltz,patkosszep}.
Within the kinetic theory we would write the current as
\be
j_{{\rm kin}}(y) = 12\lambda \phi (y) \int dP f(y,p) \, , \label{classicalcurrent}
\ee
where the integration measure $dP$ enforces positivity of the energy and
an on--shell constraint, 
\be
dP = \frac{ d^4 \! p}{(2\pi)^3} \, 2\Theta (p_0) \delta (p_0^2 - \vec{p}^{\, 2}) \, ,
\ee
and $f(y,p)$ is a solution to the collisionless Boltzmann equation,
\be
\left ( \frac{dy_{\mu}}{d\tau} \frac{\partial}{\partial y_{\mu}} + 
\frac{dp_{\nu}}{d\tau} \frac{\partial}{\partial p_{\nu}} \right ) f(y,p) = \left ( \dot{y} \cdot \partial^{(y)}
+ \dot{p} \cdot \partial^{(p)} \right ) f(y,p) = 0 \, ,
\ee
with $\tau$ the proper time. Comparing Eq.~\pref{classicalcurrent} with the world line
formulation, Eq.~\pref{currentstart}, we then have to identify
\be
\int dP f(y,p) &=& \langle \, \int d\tau \delta^4 (y-x(\tau)) \, \rangle 
= 2\sum_{n=1}^{\infty} \int_0^{\infty} \frac{dt}{t} \int {\cal D} \! p e^{-\int_0^t dt_1 p^2} \times \nonumber \\
&& \int_{P_T} {\cal D}
\! x e^{-\int_0^t dt_1 \left ( \dot{x}^2/4 + 12\lambda \phi^2_s (x(t_1)) \right) } \left [ \int_0^t dt_2
\delta^4 (y-x(t_2)) \right ] \, , \nonumber \\  \label{qandccorr}
\ee
where $P_T$ is the boundary condition given in Eq.~\pref{theboundcond}.

In the weak coupling limit, $f(y,p)$ can be expanded in a power series in the coupling constant,
$\lambda$,
\be
f(y,p) = f_0 (y,p) + \lambda f_1 (y,p) + {\rm O} (\lambda^2) \, .
\ee
To lowest order, the kinetic theory then gives 
\be
j^{(1)}_{{\rm kin}}(y) = 12\lambda \phi (y) \int dP f_0 (y,p) \, .
\ee
With $f_0$ equal to the equilibrium Bose--Einstein distribution $n_{{\rm B}} = (e^{\beta p_0}-1)^{-1}$, 
this reproduces the lowest order correction from the one--loop effective action,
Eq.~\pref{finalcurrent}, as expected.

We can also compare the next correction (Fig. 1b with four soft fields), 
in the limit of zero external four--momentum. Collecting the same powers of $\lambda$,
by using $\dot{p}_{\mu} \sim \lambda \partial_{\mu} \phi_s^2$ (which also follows from the world line action),
the Boltzmann equation gives the solution for $f_1(y,p)$ as
\cite{phi4boltz, patkosszep, fromprop}, 
\be
\int dP f_1 (y,p) \sim \phi^2 (y) \int dP \frac{1}{p_0} \frac{df_0}{dp_0} \sim
\phi^2 (y) \frac{T}{\Lambda} \, ,
\ee
and the corresponding term for the current becomes,
\be 
j^{(2)}_{{\rm kin}}(y) \propto \lambda^2 \phi^3 (y) \frac{T}{\Lambda} \, , \label{classicalwithmu}
\ee
where $\Lambda$ is the infrared cutoff on the momentum of the hard particles.

In the world line formulation, instead of a direct infrared cutoff 
we will assume a mass $m$ for the scalar field. Neglecting any variation of the external soft field,
the one--loop effective action gives a contribution
\begin{equation} 
j^{(2)}(y) = -\frac{18\lambda^2 \phi^3 (y)}{\pi^2} \sum_{n=1}^{\infty} \int_0^{\infty} \frac{dt}{t} 
e^{-m^2t-n^2 \beta^2 /4t} = -\frac{36\lambda^2 \phi^3 (y)}{\pi^2} \sum_{n=1}^{\infty} K_0 (m\beta n)
\nonumber \\ 
\end{equation}

\vspace{2mm}
\noindent
with $K_0 (m\beta n)$ the modified Bessel function of the second kind. 
For $m\beta \rightarrow 0$, the leading term comes from
\be 
\sum_{n=1}^{\infty} K_0 (m\beta n) = \frac{\pi T}{2m} + {\rm O} (\log [ m\beta] ) \, ,
\ee
which gives a current in agreement with Eq.~\pref{classicalwithmu}, provided we take $m \sim \Lambda$. 

This agreement between the two different methods indicates that the
conjectured identity in Eq.~\pref{qandccorr} indeed is true, and gives
further evidence for the close relationship between the classical
kinetic theory and the one--loop effective action. In particular, 
Eq.~\pref{qandccorr} demonstrates how the thermal average of the microscopic one--particle
distribution, $\int d\tau \delta^4 (y-x(\tau))$, naturally appears in the one--loop effective action,
when written as a point particle path integral. It also incorporates all the contributions from the
soft field, which can then be expanded in a power series in $\lambda$, just like $f(y,p)$. Thus,
it provides an explanation for the apparent
success of using a classical transport theory for the hard particles
to obtain the effective action for the soft modes.

\end{document}